\documentclass[superscriptaddress,twocolumn,twoside,prd,floatfix,letterpaper]{revtex4}

\usepackage{graphicx}
\usepackage{amsmath}
\usepackage{amssymb}
\usepackage{fancyhdr}
\usepackage{rotating}

\def \Vista {{\sc Vista}}
\def \Sleuth {{\sc Sleuth}}

\def \ptmiss {{\ensuremath{\,/\!\!\!p_{T}}}}
\def \SumPt {{\ensuremath{\sum{p_T}}}}

\def \scriptP {\ensuremath{{\cal P}}}
\def \tildeScriptP {\ensuremath{\tilde{\cal P}}}

\def \lumi {2.0~fb$^{-1}$}

\begin{document}

\title{Global Search for New Physics with 2.0~fb$^{-1}$ at CDF}



\affiliation{Institute of Physics, Academia Sinica, Taipei, Taiwan 11529, Republic of China} 
\affiliation{Argonne National Laboratory, Argonne, Illinois 60439} 
\affiliation{University of Athens, 157 71 Athens, Greece} 
\affiliation{Institut de Fisica d'Altes Energies, Universitat Autonoma de Barcelona, E-08193, Bellaterra (Barcelona), Spain} 
\affiliation{Baylor University, Waco, Texas  76798} 
\affiliation{Istituto Nazionale di Fisica Nucleare Bologna, $^w$University of Bologna, I-40127 Bologna, Italy} 
\affiliation{Brandeis University, Waltham, Massachusetts 02254} 
\affiliation{University of California, Davis, Davis, California  95616} 
\affiliation{University of California, Los Angeles, Los Angeles, California  90024} 
\affiliation{University of California, San Diego, La Jolla, California  92093} 
\affiliation{University of California, Santa Barbara, Santa Barbara, California 93106} 
\affiliation{Instituto de Fisica de Cantabria, CSIC-University of Cantabria, 39005 Santander, Spain} 
\affiliation{Carnegie Mellon University, Pittsburgh, PA  15213} 
\affiliation{Enrico Fermi Institute, University of Chicago, Chicago, Illinois 60637} 
\affiliation{Comenius University, 842 48 Bratislava, Slovakia; Institute of Experimental Physics, 040 01 Kosice, Slovakia} 
\affiliation{Joint Institute for Nuclear Research, RU-141980 Dubna, Russia} 
\affiliation{Duke University, Durham, North Carolina  27708} 
\affiliation{Fermi National Accelerator Laboratory, Batavia, Illinois 60510} 
\affiliation{University of Florida, Gainesville, Florida  32611} 
\affiliation{Laboratori Nazionali di Frascati, Istituto Nazionale di Fisica Nucleare, I-00044 Frascati, Italy} 
\affiliation{University of Geneva, CH-1211 Geneva 4, Switzerland} 
\affiliation{Glasgow University, Glasgow G12 8QQ, United Kingdom} 
\affiliation{Harvard University, Cambridge, Massachusetts 02138} 
\affiliation{Division of High Energy Physics, Department of Physics, University of Helsinki and Helsinki Institute of Physics, FIN-00014, Helsinki, Finland} 
\affiliation{University of Illinois, Urbana, Illinois 61801} 
\affiliation{The Johns Hopkins University, Baltimore, Maryland 21218} 
\affiliation{Institut f\"{u}r Experimentelle Kernphysik, Universit\"{a}t Karlsruhe, 76128 Karlsruhe, Germany} 
\affiliation{Center for High Energy Physics: Kyungpook National University, Daegu 702-701, Korea; Seoul National University, Seoul 151-742, Korea; Sungkyunkwan University, Suwon 440-746, Korea; Korea Institute of Science and Technology Information, Daejeon, 305-806, Korea; Chonnam National University, Gwangju, 500-757, Korea} 
\affiliation{Ernest Orlando Lawrence Berkeley National Laboratory, Berkeley, California 94720} 
\affiliation{University of Liverpool, Liverpool L69 7ZE, United Kingdom} 
\affiliation{University College London, London WC1E 6BT, United Kingdom} 
\affiliation{Centro de Investigaciones Energeticas Medioambientales y Tecnologicas, E-28040 Madrid, Spain} 
\affiliation{Massachusetts Institute of Technology, Cambridge, Massachusetts  02139} 
\affiliation{Institute of Particle Physics: McGill University, Montr\'{e}al, Canada H3A~2T8; and University of Toronto, Toronto, Canada M5S~1A7} 
\affiliation{University of Michigan, Ann Arbor, Michigan 48109} 
\affiliation{Michigan State University, East Lansing, Michigan  48824}
\affiliation{Institution for Theoretical and Experimental Physics, ITEP, Moscow 117259, Russia} 
\affiliation{University of New Mexico, Albuquerque, New Mexico 87131} 
\affiliation{Northwestern University, Evanston, Illinois  60208} 
\affiliation{The Ohio State University, Columbus, Ohio  43210} 
\affiliation{Okayama University, Okayama 700-8530, Japan} 
\affiliation{Osaka City University, Osaka 588, Japan} 
\affiliation{University of Oxford, Oxford OX1 3RH, United Kingdom} 
\affiliation{Istituto Nazionale di Fisica Nucleare, Sezione di Padova-Trento, $^x$University of Padova, I-35131 Padova, Italy} 
\affiliation{LPNHE, Universite Pierre et Marie Curie/IN2P3-CNRS, UMR7585, Paris, F-75252 France} 
\affiliation{University of Pennsylvania, Philadelphia, Pennsylvania 19104}
\affiliation{Istituto Nazionale di Fisica Nucleare Pisa, $^y$University of Pisa, $^z$University of Siena and $^{aa}$Scuola Normale Superiore, I-56127 Pisa, Italy} 
\affiliation{University of Pittsburgh, Pittsburgh, Pennsylvania 15260} 
\affiliation{Purdue University, West Lafayette, Indiana 47907} 
\affiliation{University of Rochester, Rochester, New York 14627} 
\affiliation{The Rockefeller University, New York, New York 10021} 
\affiliation{Istituto Nazionale di Fisica Nucleare, Sezione di Roma 1, $^{bb}$Sapienza Universit\`{a} di Roma, I-00185 Roma, Italy} 

\affiliation{Rutgers University, Piscataway, New Jersey 08855} 
\affiliation{Texas A\&M University, College Station, Texas 77843} 
\affiliation{Istituto Nazionale di Fisica Nucleare Trieste/Udine, $^{cc}$University of Trieste/Udine, Italy} 
\affiliation{University of Tsukuba, Tsukuba, Ibaraki 305, Japan} 
\affiliation{Tufts University, Medford, Massachusetts 02155} 
\affiliation{Waseda University, Tokyo 169, Japan} 
\affiliation{Wayne State University, Detroit, Michigan  48201} 
\affiliation{University of Wisconsin, Madison, Wisconsin 53706} 
\affiliation{Yale University, New Haven, Connecticut 06520} 
\author{T.~Aaltonen}
\affiliation{Division of High Energy Physics, Department of Physics, University of Helsinki and Helsinki Institute of Physics, FIN-00014, Helsinki, Finland}
\author{J.~Adelman}
\affiliation{Enrico Fermi Institute, University of Chicago, Chicago, Illinois 60637}
\author{T.~Akimoto}
\affiliation{University of Tsukuba, Tsukuba, Ibaraki 305, Japan}
\author{M.G.~Albrow}
\affiliation{Fermi National Accelerator Laboratory, Batavia, Illinois 60510}
\author{B.~\'{A}lvarez~Gonz\'{a}lez}
\affiliation{Instituto de Fisica de Cantabria, CSIC-University of Cantabria, 39005 Santander, Spain}
\author{S.~Amerio$^x$}
\affiliation{Istituto Nazionale di Fisica Nucleare, Sezione di Padova-Trento, $^x$University of Padova, I-35131 Padova, Italy} 

\author{D.~Amidei}
\affiliation{University of Michigan, Ann Arbor, Michigan 48109}
\author{A.~Anastassov}
\affiliation{Northwestern University, Evanston, Illinois  60208}
\author{A.~Annovi}
\affiliation{Laboratori Nazionali di Frascati, Istituto Nazionale di Fisica Nucleare, I-00044 Frascati, Italy}
\author{J.~Antos}
\affiliation{Comenius University, 842 48 Bratislava, Slovakia; Institute of Experimental Physics, 040 01 Kosice, Slovakia}
% ...

\collaboration{CDF Collaboration\footnote{With visitors from $^a$University of Massachusetts Amherst, Amherst, Massachusetts 01003,
$^b$Universiteit Antwerpen, B-2610 Antwerp, Belgium, 
$^c$University of Bristol, Bristol BS8 1TL, United Kingdom,
$^d$Chinese Academy of Sciences, Beijing 100864, China, 
$^e$University of California Irvine, Irvine, CA  92697, 
$^f$Cornell University, Ithaca, NY  14853, 
$^g$University of Cyprus, Nicosia CY-1678, Cyprus, 
$^h$University College Dublin, Dublin 4, Ireland,
$^i$University of Edinburgh, Edinburgh EH9 3JZ, United Kingdom, 
$^j$Universidad Iberoamericana, Mexico D.F., Mexico,
$^k$Queen Mary, University of London, London, E1 4NS, England,
$^l$University of Manchester, Manchester M13 9PL, England, 
$^m$Nagasaki Institute of Applied Science, Nagasaki, Japan, 
$^n$University of Notre Dame, Notre Dame, IN 46556,
$^o$University de Oviedo, E-33007 Oviedo, Spain, 
$^p$Simon Fraser University, Vancouver, British Columbia, Canada V6B 5K3,
$^q$Texas Tech University, Lubbock, TX  79409, 
$^r$IFIC(CSIC-Universitat de Valencia), 46071 Valencia, Spain,
$^s$University of Virginia, Charlottesville, VA  22904,
$^{dd}$On leave from J.~Stefan Institute, Ljubljana, Slovenia, 
}}
\noaffiliation

\date{\today}





\begin{abstract}
Data collected in Run II of the Fermilab Tevatron are searched for indications of new electroweak-scale physics.  Rather than focusing on particular new physics scenarios, CDF data are analyzed for discrepancies with the standard model prediction.  A model-independent approach (\Vista) considers gross features of the data, and is sensitive to new large cross-section physics.  
Further sensitivity to new physics is provided by two additional algorithms:
a Bump Hunter searches invariant mass distributions for ``bumps'' that could indicate resonant production of new particles, and the \Sleuth\ procedure scans for data excesses at large summed transverse momentum.
This combined global search for new physics in \lumi\ of $p\bar{p}$ collisions at $\sqrt{s}=1.96$~TeV reveals no indication of physics beyond the standard model.
\end{abstract}


\maketitle



The standard model (SM) of particle physics has been remarkably successful in describing observed phenomena, but is generally believed to require expansion.
Using data corresponding to an integrated luminosity of \lumi\ of $p\bar{p}$ collisions at $\sqrt{s}=1.96$~TeV collected by the CDF II detector at the Fermilab Tevatron, we present a broad search for physics beyond the standard model without focusing on any specific proposed scenario. 
A similar search has previously been performed by the CDF Collaboration with 927~pb$^{-1}$ of data~\cite{VistaCdfPrd1}.

Events containing one or more particles with large transverse momentum ($p_T$) are analyzed for discrepancies relative to the SM prediction.
A model-independent approach (\Vista) considers gross features of the data, and is sensitive to new large cross-section physics.  
Further sensitivity to beyond-SM physics is provided by two additional algorithms:
a Bump Hunter searches invariant mass distributions for ``bumps'' that could indicate resonant production of new particles, and the \Sleuth\ procedure scans for data excesses at large summed transverse momentum.
These global algorithms provide a complementary approach to searches optimized for more specific new physics scenarios.




CDF II~\cite{WandZCrossSectionPaper:Abulencia:2005ix} 
is a general-purpose detector for high-energy $p\bar{p}$ collisions.
Tracking for charged particles is provided by silicon strip detectors and a gas drift chamber inside a 1.4~T magnetic field.
The tracking system is surrounded by electromagnetic and hadronic calorimeters and enclosed by muon detectors.  


The \Vista\ procedure is extensively described in~\cite{VistaCdfPrd1}.
A standard set of object identification criteria is used to identify isolated and energetic objects produced in the hard collision, including electrons ($e^\pm$), muons ($\mu^\pm$), taus ($\tau^\pm$), photons ($\gamma$), jets ($j$), jets originating from a bottom quark ($b$), and missing transverse momentum (\ptmiss) \endnote{We use a cylindrical coordinate system where the $z$-axis is in the direction of the proton beam, and $\theta$ and $\phi$ are respectively the polar and azimuthal angles.  The pseudorapidity is defined as $\eta = -\ln(\tan(\theta/2))$. 
The transverse momentum $p_T$ is defined as $p_T = p \sin\theta$. The missing transverse momentum \ptmiss\ is defined as the magnitude of the transverse component of the negative vector sum of the four-vectors of all identified objects and unclustered momentum in an event, where unclustered momentum is visible in the detector but not clustered into an identified object.}.
All objects are required to have $p_T \ge 17$~GeV/c. With all event selections applied, over $4\times10^6$ high-$p_T$ events are analyzed in this global search.
The standard model prediction is based on Monte Carlo event generators and a simulation of the response of the CDF detector.
Data and Monte Carlo events are partitioned into exclusive final states labeled according to the number and type of objects ($e^\pm$, $\mu^\pm$, $\tau^\pm$, $\gamma$, $j$, $b$, \ptmiss) identified in each event.  

To obtain an accurate standard model prediction, a correction model is used to improve
systematic deficiencies in the Monte Carlo theoretical prediction and the simulation of the detector response -- this information can only be obtained from the data themselves.
The details of this correction model are motivated by individual discrepancies noted in a global comparison of CDF high-$p_T$ data to the SM prediction; however, the correction model is intentionally kept as simple as possible in order to avoid over-tuning.
The correction model includes specific correction factors for the integrated luminosity of the sample, the ratios ($k$-factors) of the actual cross sections for SM processes to the  leading order approximations given by event generators, object identification efficiencies, object misidentification rates, and trigger efficiencies.  
Values for the correction factors are determined from a global fit to the data: a global $\chi^2$ is formed by comparison to the SM prediction, and minimized as a function of the correction factors.  
External information (such as higher-order cross-section calculations) is used to constrain 26 of the 43 total correction factors.
A number of minor improvements have been made to the correction model since~\cite{VistaCdfPrd1}; these changes are described in detail in~\cite{georgios_thesis}.

\begin{figure}
\includegraphics[width=3.25in]{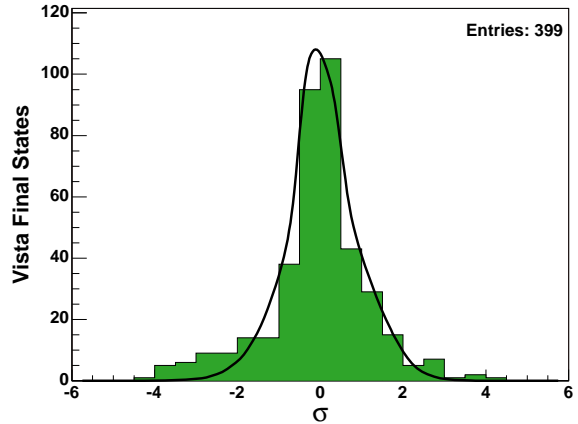}
\caption{Distribution of observed discrepancy between data and the SM prediction for populations of final states, measured in units of standard deviation ($\sigma$). The black line represents the theoretical expectation assuming no new physics.}
\label{fig:vista_summary_norm}
\end{figure}

\begin{figure}
\includegraphics[width=3.25in]{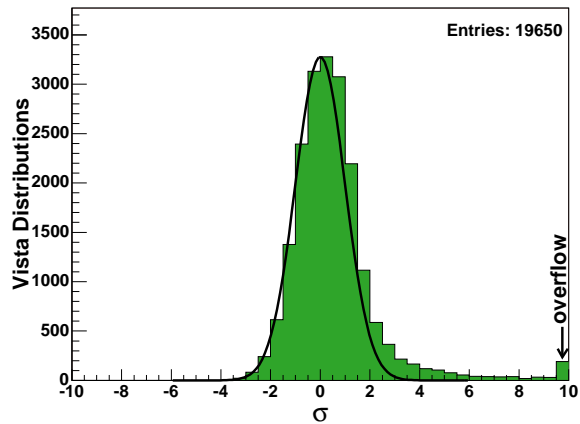}
\caption{Distribution of observed discrepancy between data and the SM prediction for shapes of kinematic distributions, measured in units of standard deviation ($\sigma$). The black line represents the theoretical expectation assuming no new physics.}
\label{fig:vista_summary_shape}
\end{figure}

The first stage in the \Vista\ global comparison is to study the populations of the exclusive final states, compared to the SM expectation.
Figure~\ref{fig:vista_summary_norm} summarizes the population discrepancies in all 399 final states, and the ten final states with the largest deviation from the SM expectation are listed in Table~\ref{table:vista_top_ten}.
After accounting for the trials factor associated with considering many final states, we find that no final state exhibits a statistically significant population discrepancy.

\begin{table}
{\mbox{\begin{tabular}{l@{ }r@{ }r@{ $\pm$ }l@{ }c@{ }c}
\hline
\hline
Final State &  Data & \multicolumn{2}{c}{Background} & $\sigma$ & $\sigma_{t}$  \\ \hline 
$b$$e^\pm$\ptmiss\ & $690$ & $817.7$ & $9.2$ & $-4.3$ & $-2.7$ \\ 
$\gamma$$\tau^\pm$ & $1371$ & $1217.6$ & $13.3$ & $+4.0$ & $+2.2$ \\ 
$\mu^\pm$$\tau^\pm$ & $63$ & $35.2$ & $2.8$ & $+3.7$ & $+1.7$ \\ 
$b$$2j$\ptmiss\ ($\Sigma p_T > 400$~GeV) & $255$ & $327.2$ & $8.9$ & $-3.7$ & $-1.7$ \\ 
$2j$$\tau^\pm$ ($\Sigma p_T < 400$~GeV) & $574$ & $670.3$ & $8.6$ & $-3.6$ & $-1.5$ \\ 
$3j$$\tau^\pm$ ($\Sigma p_T < 400$~GeV) & $148$ & $199.8$ & $5.2$ & $-3.5$ & $-1.4$ \\ 
$e^\pm$$\tau^\pm$\ptmiss & $36$ & $17.2$ & $1.7$ & $+3.5$ & $+1.4$ \\ 
$2j$$\tau^\pm$$\tau^\mp$ & $33$ & $62.1$ & $4.3$ & $-3.5$ & $-1.3$ \\ 
$e^\pm$$j$ & $741710$ & $764832$ & $6447.2$ & $-3.5$ & $-1.3$ \\ 
$j$$2\tau^\pm$ & $105$ & $150.8$ & $6.3$ & $-3.4$ & $-1.2$ \\ 
\hline
\hline
\end{tabular}
}}
\caption{The ten most discrepant \Vista\ final states, showing the number of data events observed and the number of background events expected. Only Monte Carlo statistical uncertainties on the background prediction are included. $\sigma$ and $\sigma_t$ represent the level of discrepancy, before and after accounting for the trials factor.} 
\label{table:vista_top_ten}

\end{table}

The \Vista\ global comparison also considers the shapes of kinematic distributions. The Kolmogorov-Smirnov test is used to assess the agreement between data  and the SM prediction for 19 650 distributions. The results are summarized in Fig.~\ref{fig:vista_summary_shape} which shows the degree of discrepancy measured for each distribution, displayed in units of standard deviation ($\sigma$).
Distributions exhibiting disagreement between data and SM prediction are at large positive $\sigma$.

\begin{figure}
\includegraphics[angle=-90,width=3.25in]{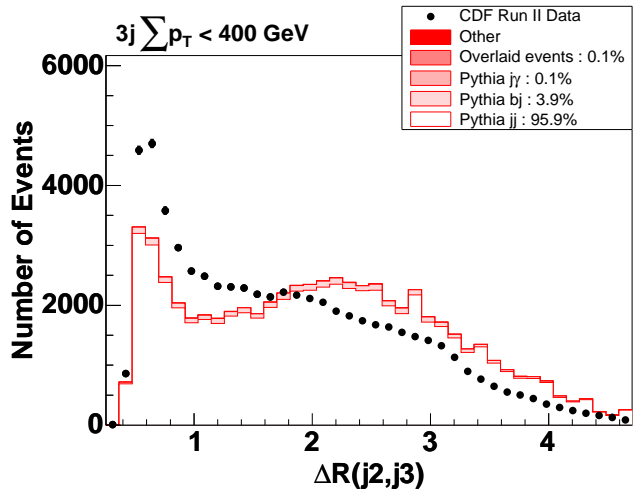}
\caption{One of the significant shape discrepancies seen by \Vista, $\Delta R = \sqrt{\Delta \phi^2 + \Delta \eta^2}$ between the second and third highest $p_T$ jets in the \Vista\ 3-jet final state. Data are shown as  filled (black) circles, with the standard model prediction shown as the shaded (red) histogram.}
\label{fig:3j_effect}
\end{figure}

We find that 555 distributions have a significant discrepancy, which is defined as being greater than 3$\sigma$ after accounting for the trials factor associated with the number of distributions considered. 
The discrepant distributions fall into three categories. Residual
``crudeness'' in the correction model, primarily from using simplified
$p_T$ dependences for fake rate correction functions, accounts for 3\%
. Another 16\% are attributed to an inadequate modeling of the
transverse boost of the colliding system.  
The remaining 81\% most likely arise from incorrect modeling of soft QCD parton showering. This is best exemplified by Fig.~\ref{fig:3j_effect}, which shows $\Delta R = \sqrt{\Delta \phi^2 + \Delta \eta^2}$ between the second and third highest $p_T$ jets in the \Vista\ 3-jet final state.  This observation has been discussed in more detail in~\cite{VistaCdfPrd1}.
The nature of these shape discrepancies does not warrant treating any of them as indicative of potential new physics.

A statistically significant local excess of data in an invariant mass
variable would be the most direct evidence of resonant production of a
new particle.  The Bump Hunter algorithm is designed to identify mass
resonances with a narrow natural width that would appear as Gaussian
bumps on top of the SM background, with width equal to the detector
resolution.   
The Bump Hunter searches in all exclusive final states, and examines all mass variables that can be constructed from combinations of the final state objects. If there is \ptmiss, transverse mass variables are also considered. The SM background is obtained from the \Vista\ procedure.

The method is described in detail in \cite{georgios_thesis}.
Each mass variable is scanned with a sliding window of width equal to twice the typical detector resolution for the component objects. 
Only windows that contain at least 5 data events are considered.
The p-value for the window is defined as the Poisson probability that the expected SM background would fluctuate up to or above the number of data events observed.
To ensure the window really represents a bump of the correct resolution-based width and not some broader excess, the ``side-bands'' of equal width on either side of the central window are required to be within 5 standard deviations of the SM expectation, and less discrepant than the central window.




In each mass variable, the bump candidate with the smallest p-value is selected. 
The significance of this bump is given by $P_a$, the fraction of pseudoexperiments which would have produced a more interesting bump in this mass variable purely by random fluctuations of the SM background. $P_a$ incorporates the trials factor associated with examining multiple overlapping windows within the mass variable.
For computational reasons, it is prohibitive to determine $P_a$ by pseudoexperiments for all mass variables, so instead an analytic approximation is used. If the analytic estimation returns a value of $P_a$ with a significance of $\ge 4.5\sigma$, then pseudoexperiments are performed for accurate determination.

Each mass variable is further assigned a probability $P_b$, defined as the probability under the null hypothesis that any mass variable would appear more significant than this. 
Assuming no correlations, $P_b = 1 - (1-P_a)^N$, where $N$ is the total number of mass variables examined.
If $P_b$ corresponds to a significance of $\ge 3\sigma$, 
that effect is then considered as potentially due to new physics.


\begin{figure}
\centering
\includegraphics[angle=0,width=0.9\columnwidth]{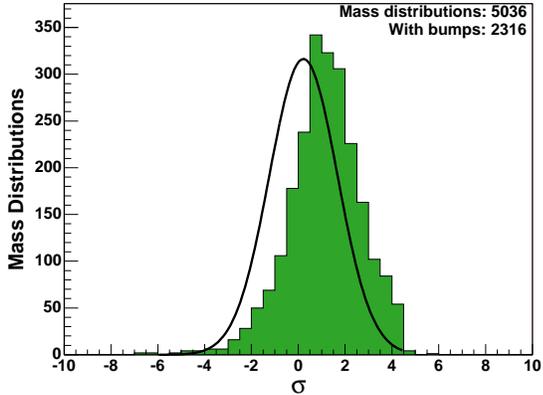}
\caption[Significance of the most interesting bump in each mass variable.]{Significance of the most interesting bump in each mass variable ($P_a$, in units of standard deviations) considered by the Bump Hunter. The black line represents the theoretical expectation assuming no new physics.
}
\label{fig:vistaSummary_bump}
\end{figure}

The Bump Hunter examines 5036 mass variables, of which 2316 are found to have at least one local excess  satisfying our bump definition (the other variables are mainly from small-population final states which fail to satisfy the criterion of 5 data events in a mass window).
The expected and observed distributions of $P_a$, converted to units of $\sigma$, are shown in Fig.~\ref{fig:vistaSummary_bump}.
The distribution of $P_a$ in the data is seen to be shifted towards
positive $\sigma$ relative to the expectation, indicating disagreement
between data and the SM prediction. This reflects the fact that the Bump Hunter algorithm is quite sensitive to local features in mass variables which can arise since the Monte Carlo-based SM background prediction does not perfectly describe the data.
The sharp drop seen in the data at $P_a = 4.5 \sigma$ results from the transition between analytic estimation and accurate determination of $P_a$.

The only mass variable with a bump which exceeds the discovery threshold is the invariant mass of all four jets in the $4j$ final state, shown in Fig.~\ref{fig:4j_sumPt0-400:mass(j1,j2,j3,j4)}. 
This mass variable has $P_a$ corresponding to $5.7 \sigma$, and $P_b$ to $4.1 \sigma$.
However, this bump is attributed to the aforementioned difficulty modeling soft QCD jets and is not thought to indicate new physics.

\begin{figure}
\centering
\includegraphics[angle=-90,width=0.9\columnwidth]{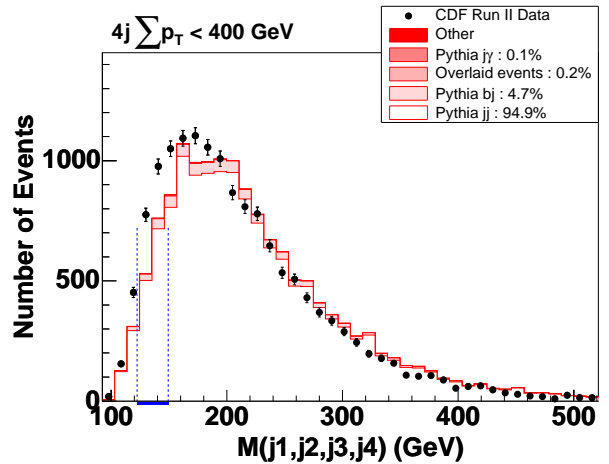}
\caption{Distribution of the invariant mass of all four jets in the $4j\;\SumPt < 400$ GeV final state. This variable contains the most significant bump found by the Bump Hunter, indicated by the dashed (blue) lines.}
\label{fig:4j_sumPt0-400:mass(j1,j2,j3,j4)}
\end{figure}




The final component of this global search for physics beyond the standard model is a procedure called \Sleuth~\cite{SleuthPRL:Abbott:2001ke,SleuthPRD1:Abbott:2000fb,SleuthPRD2:Abbott:2000gx}.
\Sleuth\ is a quasi--model-independent search technique, based on the assumption that new electroweak-scale physics will manifest itself as a high-$p_T$ excess of data over the SM expectation in a particular final state.  Tests have shown \Sleuth\ to have sensitivity comparable to targeted searches  for phenomena that satisfy \Sleuth's basic assumptions.

The procedure is identical to that used in~\cite{VistaCdfPrd1}.
The algorithm considers a single variable, the summed scalar transverse momentum (\SumPt) of all objects in the event. 
The SM prediction for the distribution of \SumPt\ is determined as part of the \Vista\ procedure. 
The exclusive final states examined by \Sleuth\ are created by merging \Vista\ final states according to certain rules described in~\cite{VistaCdfPrd1}. 
For each final state, \Sleuth\ determines the region (defined as an interval in \SumPt\ extending from a data-point up to infinity) which has the smallest probability that the SM prediction would fluctuate up to or above the number of observed data events.
The algorithm then finds \scriptP, the fraction of pseudoexperiments drawn from the SM \SumPt\ distribution which produce any region more interesting than the region found in the data.
\Sleuth\ selects the final state with the smallest value of \scriptP, and calculates the overall significance, \tildeScriptP, which accounts for the number of final states considered.
With an accurate correction model and in the absence of new physics, the distribution of \tildeScriptP\ is uniform between zero and unity; in the presence of new physics, a small value of \tildeScriptP\ is expected.  The threshold for pursuit of a possible discovery case is taken to be $\tildeScriptP < 0.001$.



\begin{figure}[tb]
\includegraphics[width=2.1in,angle=270]{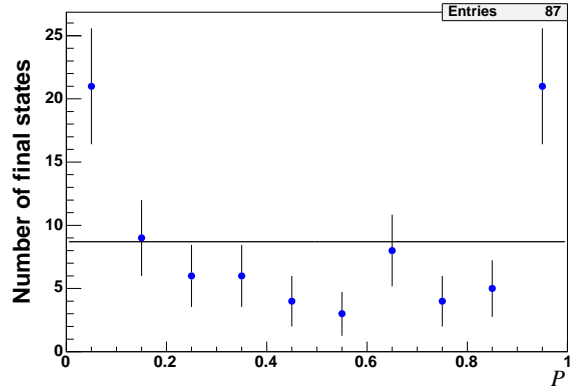}
\caption{The distribution of \scriptP\ in the data, with one entry for each final state considered by \Sleuth. The black line represents the theoretical expectation assuming no new physics.}
\label{fig:scriptPsPlots}
\end{figure}

The distribution of \scriptP\ for the final states considered by \Sleuth\ in the data is shown in Fig.~\ref{fig:scriptPsPlots}.  
The concavity of this distribution reflects the crudeness (i.e.\ under-tuning) of our correction model. A crude correction model results in more outliers than expected, which, when converted into values of \scriptP, produces excesses at the extremes of both low and high probability.

\begin{figure*}[tb]
\includegraphics[angle=-90,width=\columnwidth]{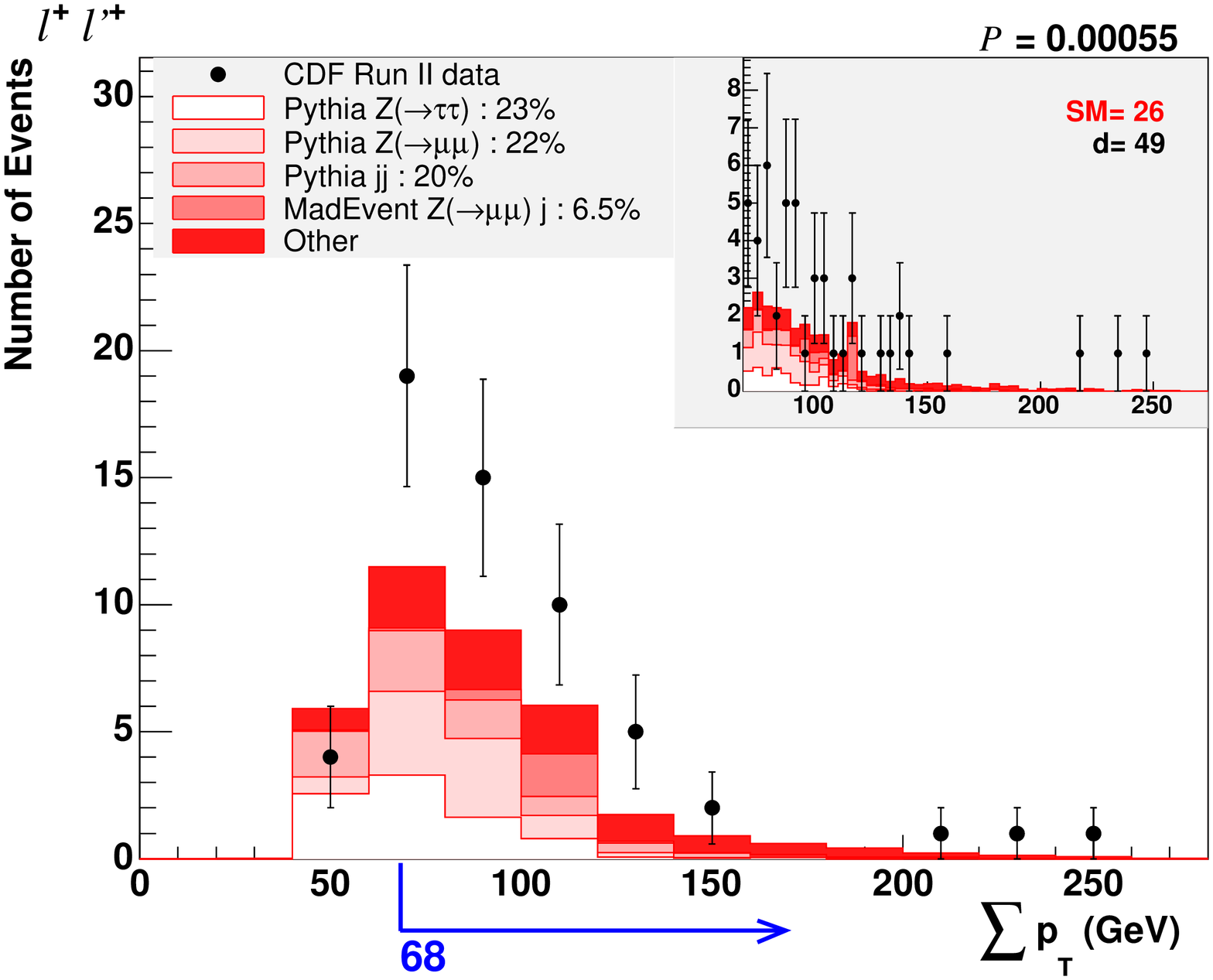}
\hfill
\includegraphics[angle=-90,width=\columnwidth]{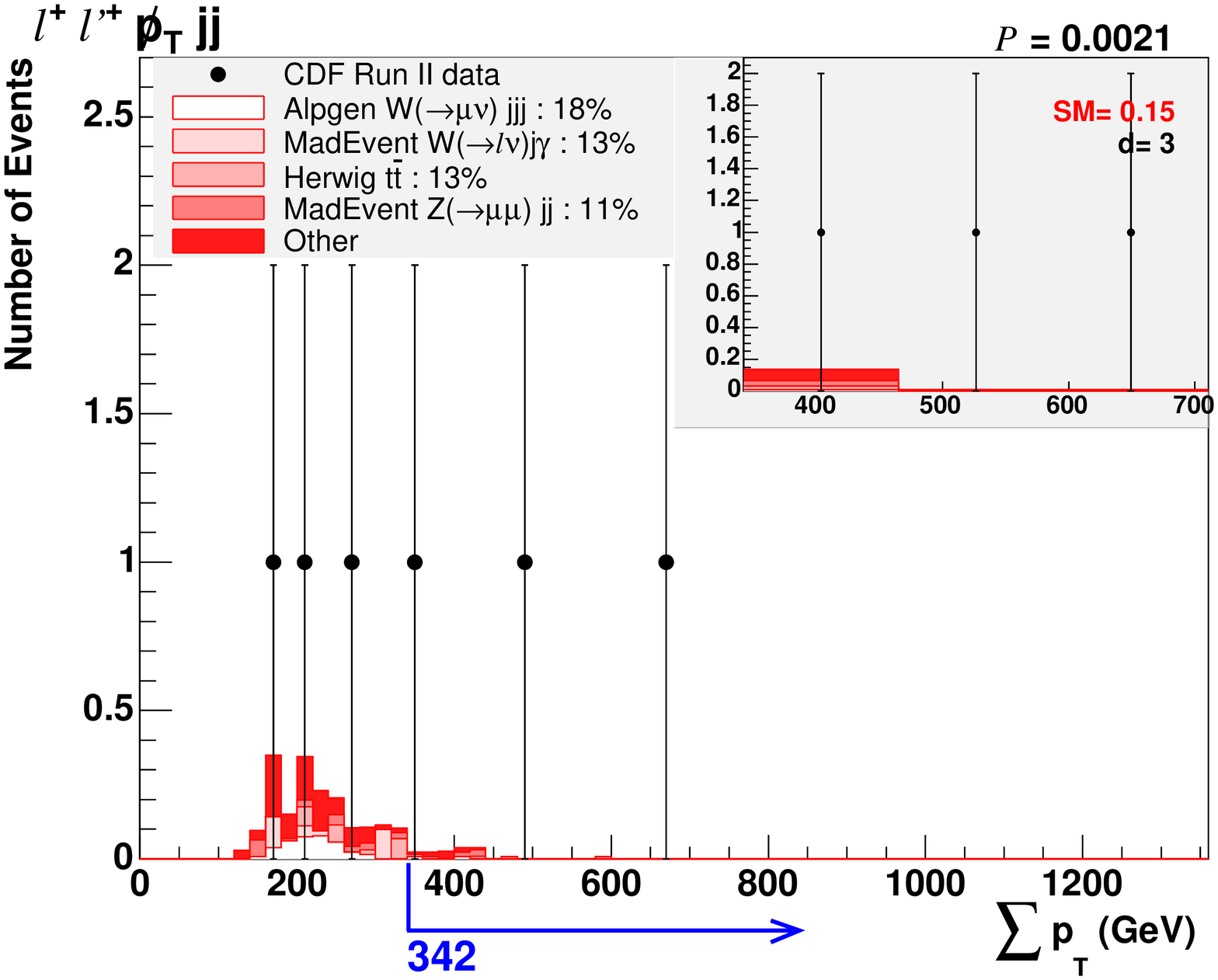}
\includegraphics[angle=-90,width=\columnwidth]{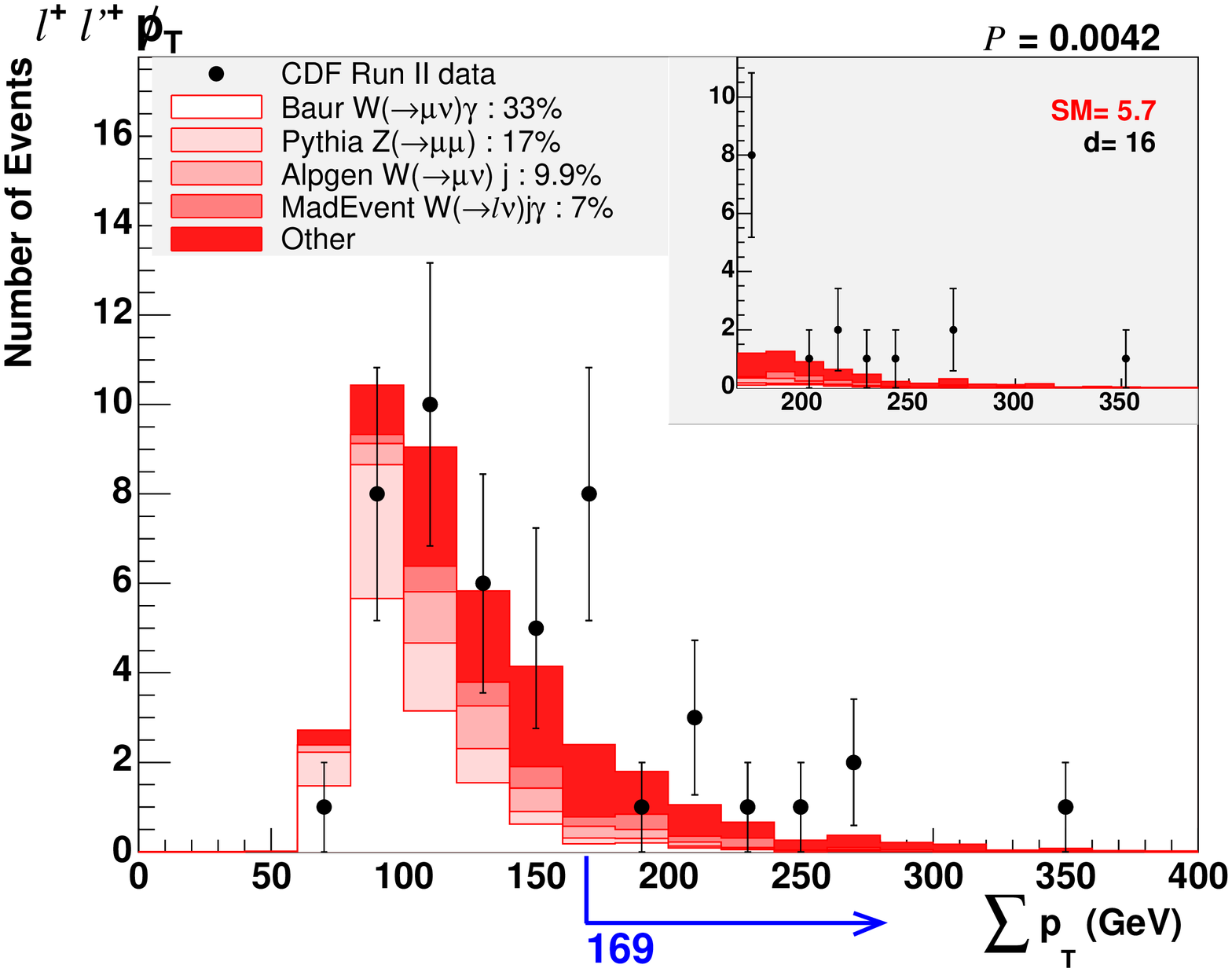}
\hfill
\includegraphics[angle=-90,width=\columnwidth]{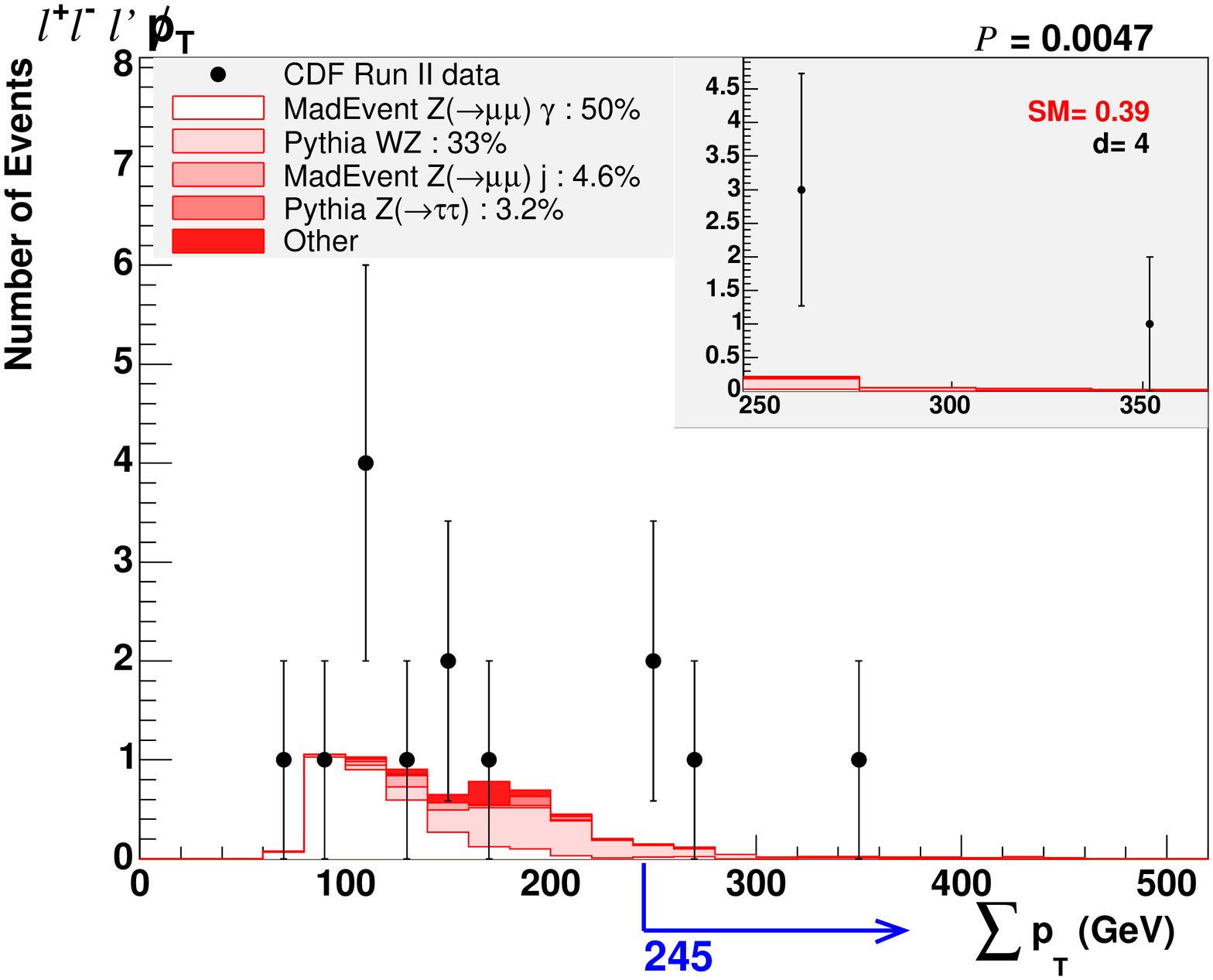}
\caption{The \SumPt\ distributions of the four most interesting final states found by \Sleuth.
Data are shown as  filled (black) circles, with the expected contribution from standard model processes shown as the shaded (red) histograms and identified in the legend. The category ``Other'' represents the sum of all remaining relevant SM processes, each of which individually is smaller than the smallest itemized contribution.
The label in the top left corner of each plot lists the objects in the final state, where {\em l}$^\pm$ is a lepton ($e$ or $\mu$), {\em l'} is an additional lepton of different flavor, $j$ denotes a jet, and ${\not\!\!p_T}$ represents missing transverse momentum. Global charge conjugation is implied, so that a final state labeled {\em l}$^+$ {\em l'}$^+$ also includes {\em l}$^-$ {\em l'}$^-$.
The region with the most significant excess of data over the SM expectation is indicated by the arrow below the $x$-axis, and displayed in the inset with the number of events expected (SM) and observed (d). The significance of the excess is shown by the value of \scriptP\ in the top right corner.}
\label{fig:sleuth}
\end{figure*}



The \SumPt\ distributions of the four most interesting final states found by \Sleuth\ are shown in Fig.~\ref{fig:sleuth}.
These are: 
$e^\pm \mu^\pm$,
$e^\pm \mu^\pm jj \ptmiss$,
$e^\pm \mu^\pm \ptmiss$, 
and $e^\pm e^\mp \mu^\pm \ptmiss + \mu^\pm \mu^\mp e^\pm \ptmiss$.
It is intriguing to note that all four contain the rare signature of a same-sign electron-muon pair. Such a signature can arise in a number of ways.
SM processes that produce real electrons and muons with the same charge include $WZ$ production with leptonic decays, where one of the leptons is not reconstructed in the detector. 
There are also processes which produce real electrons in the forward region of the CDF II detector, where the reduced tracking coverage means the electron charge sign has a higher probability of being falsely reconstructed; such processes include $t\bar{t}$ production, and $Z\rightarrow \tau^+ \tau^-$ where both taus decay leptonically.
In addition, there are processes with a real muon and a fake electron. These are largely $W/Z$+jets production, where a primary quark or gluon jet is misidentified as an electron in the detector, and $W\gamma$/$Z\gamma$, where the photon undergoes conversion to produce an electron.
Also relevant is the case when both the electron and the muon are fakes, predominantly from dijet events.
The relative proportion of these potential backgrounds varies for each final state, depending on the presence of \ptmiss\ and the number of jets.
Since all of these processes and detector effects also contribute to other more highly-populated final states where good agreement is seen, 
their rates are quite well constrained by this global analysis.

However, while it is noteworthy that the top four final states all contain the same rare signature, this is an \emph{a posteriori} observation and its significance is therefore difficult to estimate. 
\Sleuth's \emph{a priori} procedure is to calculate the significance of only the single most discrepant final state. We find that $\tildeScriptP = 0.08$, i.e.\ that $8\%$ of pseudoexperiments drawn from the \Vista\ SM implementation would have produced a more significant excess in a single final state purely by chance fluctuations. This is far from the threshold of $\tildeScriptP < 0.001$, and therefore we do not pursue this as a potential discovery.



In summary, CDF has performed a model-independent global search for new high-$p_T$ physics in \lumi of $p\bar{p}$ collisions at $\sqrt{s}=1.96$~TeV.
The populations of 399 exclusive final states are compared to a standard model prediction, but no significant discrepancy is found after accounting for the trials factor associated with looking in many places. The shapes of 19 650 kinematic distributions are also studied, and although 555 show a significant discrepancy, most of these are attributed to inadequate modeling of soft QCD jet emission in the underlying Monte Carlo prediction, rather than a sign of new physics. A Bump Hunter algorithm scans invariant mass distributions for narrow bumps that could indicate resonant production of new particles: only one significant bump is found, and it is attributed to the same underlying problem as above.
The \Sleuth\ algorithm searches the \SumPt\ spectrum of each final
state, but finds no significant excesses of data over the SM prediction in the tails of any single distribution.
This CDF global search has not discovered new physics in \lumi.



We thank the Fermilab staff and the technical staffs of the
participating institutions for their vital contributions. This work
was supported by the U.S. Department of Energy and National Science
Foundation; the Italian Istituto Nazionale di Fisica Nucleare; the
Ministry of Education, Culture, Sports, Science and Technology of
Japan; the Natural Sciences and Engineering Research Council of
Canada; the National Science Council of the Republic of China; the
Swiss National Science Foundation; the A.P. Sloan Foundation; the
Bundesministerium f\"ur Bildung und Forschung, Germany; the Korean
Science and Engineering Foundation and the Korean Research Foundation;
the Science and Technology Facilities Council and the Royal Society,
UK; the Institut National de Physique Nucleaire et Physique des
Particules/CNRS; the Russian Foundation for Basic Research; the
Ministerio de Educaci\'{o}n y Ciencia and Programa Consolider-Ingenio
2010, Spain; the Slovak R\&D Agency; and the Academy of Finland.


\bibliography{prd2_rc}

\end{document}